\documentclass[letterpaper, 10 pt, conference]{ieeeconf}  
\usepackage[utf8]{inputenc}
\usepackage[T1]{fontenc}
\usepackage{gensymb}
\usepackage{siunitx}
\IEEEoverridecommandlockouts                              

\usepackage{amsmath} 
\usepackage{amssymb}  
\usepackage{graphicx} 

\title{\LARGE \bf
Explicit MPC for Parameter Dependent Linear Systems
}

\author{Carlos J. G. Rojas$^{1}$, Esteban Lage Cano$^{1}$, Leyla Özkan$^{1}$
\thanks{*This work was not supported by any organization}
\thanks{The first two authors contributed equally to this work. $^{1}$Control Systems Group, Department of Electrical Engineering, Eindhoven University of Technology, 5612 AZ, Eindhoven, The Netherlands.
        {\tt\small c.j.gonzalez.rojas@tue.nl, e.c.lage.cano@tue.nl, l.ozkan@tue.nl.}}%
}

\begin{document}

\maketitle
\thispagestyle{empty}
\pagestyle{empty}

\begin{abstract}
This paper presents two explicit Model Predictive Control (MPC) formulations for linear systems parameterized in terms of design variables. Such parameter-dependent behavior commonly arises from operating-point–dependent linearization of nonlinear systems as well as from variations in mechanical, electrical, or thermal properties associated with material selection in the design of the process or system components. In contrast to explicit MPC approaches that treat design parameter variations and dependencies as disturbances, the proposed methods incorporate the parameters directly into the system matrices in an affine manner. However, explicitly incorporating these dependencies significantly increases the complexity of explicit MPC (eMPC) formulations due to  resulting nonlinear terms involving decision variables and parameters. We address this complexity by proposing two approximation methods. Both methods are applied on two examples and their performances are compared with respect to the exact eMPC implementation.

\end{abstract}

\section{INTRODUCTION}
Many control systems exhibit dynamics that vary with operating conditions or intrinsic properties of the physical components.  Such variations occur naturally when nonlinear systems are linearized around different operating points, as well as from changes in mechanical, electrical or thermal  characteristics associated with material selection and process design.  Regardless of their role, 
these parameters enter the model as constants within the time domain,
thus shaping the system behavior without dynamically evolving alongside the state.  When these properties are design variables rather than fixed quantities, as in control co-design problems, the resulting closed-loop behavior depends explicitly on parameters chosen for the system design stage.

The effect of parametric variations on the system can be either taken into account directly in the representation of the dynamics or considered as uncertainties/disturbances~\cite{diangelakis2017}. In most cases, control strategies, including MPC approaches, treat parameter variations as disturbances and consider robust worst-case formulations \cite{Saltik2018}. While such approaches provide robustness, they lead to conservative control actions because they do not use specific information related to the measured parameters or design/material properties. On the other hand,  explicitly incorporating parameter dependence into the prediction model allows the controller to account for the influence of operating conditions, material properties, and design choices on system behavior. This  is particularly relevant in control co-design problems, where control performance and feasibility depend jointly on the controller and on the system design parameters.

Explicit MPC has been an attractive strategy for low-order systems with real-time execution limitations since the optimization problem is solved offline and the resulting control law is implemented online as a piecewise-affine (PWA) function of the system state. Moreover, it has also been investigated in the context of integrated scheduling and control \cite{Zhuge2014}, and integration of design and control formulations \cite{diangelakis2017}. However, formulating an eMPC with parameter-dependent system matrices results in nonlinear coupling between control decision variables and parameters, which makes the resulting optimization problem difficult to solve. In order to address this problem, we propose two approximation methods. In the first method, we utilize McCormick envelopes constraints, whereas in the second method, we apply the Fréchet derivative to the matrix powers in the condensed model. The effectiveness of the proposed methods is demonstrated through two illustrative examples. These examples highlight the trade-offs between approximation accuracy and implementation complexity.

This article is organized as follows: Section \ref{sec:Preliminaries} introduces the assumptions underlying the model and control strategy. Section \ref{sec:Methods} presents the proposed methods for approximating explicit MPC laws with affine parameter dependence. Section \ref{sec:Cases} illustrates the approach through representative case studies. Section \ref{sec:Results} reports numerical results, and Section \ref{sec:Conclusion} concludes the paper.

\section{PRELIMINARIES}
\label{sec:Preliminaries}
\subsection{Systems with parametric dependence}
\label{Parametric representation}

We consider a discrete linear parameter-dependent model representation of the system,
\begin{equation}
\label{GeneralSysDef}
    {x}_{k+1} = A(\theta)\,x_k + B(\theta)\,u_k, 
\end{equation}
\begin{equation}
    y_k = C(\theta)\,x_k + D(\theta)u_k,
\end{equation}
where $x_k \in\mathbb{R}^{n}$ denotes the state at step  $k$, and the state-space matrices depend on the parameter vector $\theta$.
The dependence of the matrices on $\theta$ may be obtained by:
\begin{enumerate}
    \item \emph{Local linearization} of nonlinear dynamics around an operating point $(x^\star,u^\star,\theta)$, yielding
    \begin{equation}
        A(\theta)=\left.\frac{\partial f}{\partial x}\right|_{(x^\star,u^\star,\theta)}, 
        \qquad 
        B(\theta)=\left.\frac{\partial f}{\partial u}\right|_{(x^\star,u^\star,\theta)},
    \end{equation}
    and analogously for $C(\theta)$ and $D(\theta)$ \cite{toth2010book}.
    \item \emph{Multi-model identification}, in which local LTI models are estimated at sampled parameter values $\{\theta_k\}$ and interpolated to produce a continuous mapping $\theta \mapsto (A,B,C,D)$ \cite{toth2010book}.
\end{enumerate}
The parameter dependence of the system is captured through an affine parameterization as in the following.
\begin{equation}
    A(\theta) = A_0 + \sum_{j=1}^{n_\theta} \theta_j A_j ,
\end{equation}
where $A_0, A_j \in \mathbb{R}^{n_x \times n_x}$ are estimated according to one of the previous methods. This representation allows for the entire admissible design space to be explored using feasible values of $\theta$. In contrast, the standard model for linear eMPC solutions \cite{diangelakis2017} uses the following structure: 
\vspace{-0.4em}
\begin{equation}
    x_{k+1}= Ax_k + Bu_k+E\,d_k,
\label{EqDist}
\end{equation}
\begin{equation}
   y_k =Cx_k+Du_k,
\end{equation}
 where $E\in\mathbb{R}^{n\times r}$ maps the effects of a constant input $d_k = d$ (design parameters) on the dynamics of the system.
For simplicity, we provide the following two definitions to express the state trajectories as affine functions of the initial condition and the input sequence. 

\textbf{ Definition 1 (Condensed model for parameter-dependent representation)}
Consider the state evolution in \eqref{GeneralSysDef}, and let $\textbf{X}$ and $\textbf{U}$ denote the stacked state and input trajectories over the prediction horizon, i.e.,
\[
\textbf{X} = \begin{bmatrix} x_1^\top & \cdots & x_N^\top \end{bmatrix}^\top,
\qquad
\textbf{U} = \begin{bmatrix} u_0^\top & \cdots & u_{N-1}^\top \end{bmatrix}^\top .
\]
\vspace{-0.4em}
The unrolled dynamics are
\begin{equation}
    \textbf{X} = S_x(\theta)x_0 + S_u(\theta) \textbf{U},
\label{CondEq}
\end{equation}
with
\vspace{-0.8em}
\begin{equation}
\bigl[S_x(\theta)\bigr]_{k} = A(\theta)^{k}, \qquad k = 0,\dots,N .
\end{equation}

\begin{equation}
\bigl[S_u(\theta)\bigr]_{k,i}
=
\begin{cases}
A(\theta)^{k-1-i} B(\theta), & i \le k-1,\\[4pt]
M_0, & i > k-1 .
\end{cases}
\end{equation}
\[
k,i = 0,\dots, N-1.
\]
where $M_0 \in \mathbb{R}^{n \times m} $ is a matrix of zeros \cite{borrelli2017predictive}.

\textbf{ Definition 2 (Condensed model for parametric state-space model through exogenous input).}
Unrolling the dynamics of the system introduced in \eqref{EqDist} leads to the following condensed form:
\begin{equation}
     \textbf{X} = S_x x_0 + S_u  \textbf{U} + S_d d,
\end{equation}
\vspace{-0.4em}
where
\vspace{-0.4em}
\begin{equation}
\bigl[S_d\bigr]_k \;=\; \sum_{i=0}^{k-1} A^{i}\,E,
\qquad k=1,\dots,N .
\end{equation}
and $\textbf{X}$, $\textbf{U}$, and $S_x$, $S_u$ as given in Definition 1.

\subsection{Explicit MPC}

For a given initial state $x_0$ and a fixed parameter $\theta$, consider the condensed form given in \eqref{CondEq}. The finite-horizon MPC problem can be formulated as follows:

\textbf{Definition 3 (Condensed MPC Problem).}
\begin{align}
\label{eq:condensed_mpc}
\min_{\mathbf{U}}  J_0(x_0, \theta, \mathbf{U} )= \min_{\mathbf{U}}  \quad &
  \textbf{X}^\top \bar{Q}  \textbf{X} + \mathbf{U} ^\top \bar{R} \mathbf{U} ,
  \\
\text{subj. to}\quad G \mathbf{U} \le w + E x_0
\end{align}
with \[
\bar{Q}=I_N \otimes Q \;\in\; \mathbb{R}^{(Nn)\times(Nn)}, \quad
\bar{R}=I_N \otimes R \in \mathbb{R}^{(Nm)\times(Nm)} .
\]
and $G, w$ and $E$ obtained from the system constraints and the condensed model \cite{borrelli2017predictive}. Notice that both the cost and constraints depend on the parameter $\theta$, but we omitted it in the notation. For simplicity, we assume that only the dynamic matrix $A$ depends on $\theta$ as follows
\begin{equation}
\label{Parametric decomposition}
    A(\theta) = \tilde{A} + \theta \Delta A.
\end{equation}
If we substitute this decomposition directly into the condensed MPC formulation,  powers of the form $A(\theta)^k$ will appear in the objective function, which implies a nonlinear dependence on $\theta$. This shows that even for this basic case, the optimization problem is generally nonconvex and could be computationally intractable due to the polynomial dependence on $\theta$ \cite{besselmann2012}.
If the explicit dependence on $\theta$ is omitted, the objective function defined in \eqref{eq:condensed_mpc} can be rewritten in quadratic form.

\textbf{Definition 4 (Multiparametric Quadratic Program).}
\begin{equation}
J_0(x_0,\mathbf{U} )= \mathbf{U}^\top H \mathbf{U} + 2  f(x_0) \mathbf{U} + c.
\end{equation}
Since the constraints of this optimization problem are affine, this is a \emph{multiparametric quadratic program} (mpQP) with 
the initial condition \(x_0\) as the parameter.
Its solution over the admissible set of initial states can be obtained using region-based or regionless approaches \cite{drgona2016regionless}.

\section{APPROXIMATION METHODS}
\label{sec:Methods}
In this section, we present two methods that reduce the complexity of employing the state-space parametric dependent representation in the eMPC formulation. The first method handles the nonlinear terms through the introduction of McCormick constraints, and the second method  employs first-order approximations through the use of the Fr\'echet derivative. Both methods consider the affine decomposition presented in~\eqref{Parametric decomposition}, where $\tilde{A}$ and $\Delta A$ could be obtained by recentering and rescaling ($\theta \in [-1, 1]$) or using the maximum and minimum values considered for the system ($\theta \in [0, 1]$).

\subsection{Method I: Explicit MPC for Parameter-Dependent Systems Using McCormick Relaxations}

We construct a first-order approximation of the lifted prediction model of the form
\begin{equation}
X \approx S_{x_0} x_0 + S_u U
+ \sum_{p=1}^{n_\theta} S_{\theta_p x_0} (\theta_p x_0)
+ \sum_{p=1}^{n_\theta} S_{\theta_p u} (\theta_p U),
\label{eq:approx_lifted_model_multi}
\end{equation}

where $S_{x_0}$ and $S_u$ denote the standard prediction matrices associated with the nominal system matrices
$\tilde A$ and $B$, respectively.
The matrices $S_{\theta_p x_0}$ and $S_{\theta_p u}$ capture the sensitivity of the predicted state trajectory with respect to the parameter component $\theta_p$.

The contribution of the initial condition $x_0$ through the parameter $\theta_p$ to the predicted state at step $i$ is given by the block
$S_{\theta_p x_0}^{(i)} \in \mathbb{R}^{n_x \times n_x}$,
\begin{equation}
S_{\theta_p x_0}^{(i)} =
\sum_{\ell=0}^{i-1} \tilde{A}^{\,i-1-\ell} \Delta A_p \tilde{A}^{\,\ell},
\qquad i = 1, \dots, N,
\label{eq:Ctheta_x0_multi}
\end{equation}
where $\Delta A_p$ denotes the matrix associated with the $p$-th parameter.
Stacking these blocks vertically yields
$S_{\theta_p x_0} \in \mathbb{R}^{N n_x \times n_x}$.

Similarly, the contribution of the bilinear term $\theta_p u_j$ to the predicted state $x_i$ is approximated by
\begin{equation}
S_{\theta_p u}^{(i,j)} =
\sum_{\ell=0}^{i-1-j} \tilde{A}^{\,i-1-j-\ell} \Delta A_p \tilde{A}^{\,\ell}
+ \tilde{A}^{\,i-1-j} B,
\qquad j < i,
\label{eq:Ctheta_u_multi}
\end{equation}
and is zero otherwise.
Collecting all such blocks yields the lower block-triangular matrix
$S_{\theta_p u} \in \mathbb{R}^{N n_x \times N n_u}$.

The lifted model~\eqref{eq:approx_lifted_model_multi} contains bilinear terms involving parameters and decision variables.
To render the formulation compatible with standard explicit MPC techniques, these bilinearities are treated as follows. First, the product $\theta_p x_0$ involves parameters only and is therefore introduced as an auxiliary parameter
\[
\chi_p := \theta_p x_0 \in \mathbb{R}^{n_x},
\]
which augments the parameter vector by $n_x$ components for each $\theta_p$. Second, the product $\theta_p U$ involves both a parameter and a decision variable.
To handle this term, an auxiliary input variable
\[
V_p := \theta_p U \in \mathbb{R}^{N n_u}
\]
is introduced.
The equality is enforced through McCormick envelope constraints, which provide a convex relaxation of the bilinear mapping over bounded domains.
This relaxation is exact at the boundary of the feasible set and corresponds to the convex hull of the bilinear term.

\subsection*{McCormick Envelope Constraints}

Consider the bilinear terms $u_{\theta_p,k} = \theta_p u_k$, where
$\theta_p \in [\underline{\theta}_p, \overline{\theta}_p]$ and
$u_k \in [\underline{u}, \overline{u}]$.
Each term is convexified using McCormick envelope constraints \cite{mccormick1976}, imposed elementwise over the prediction horizon,
\begin{equation}\label{eq:mccormick}
\begin{aligned}
u_{\theta_p,k} &\ge \underline{\theta}_p u_k + \theta_p \underline{u} - \underline{\theta}_p \underline{u},\\
u_{\theta_p,k} &\ge \overline{\theta}_p u_k + \theta_p \overline{u} - \overline{\theta}_p \overline{u},\\
u_{\theta_p,k} &\le \overline{\theta}_p u_k + \theta_p \underline{u} - \overline{\theta}_p \underline{u},\\
u_{\theta_p,k} &\le \underline{\theta}_p u_k + \theta_p \overline{u} - \underline{\theta}_p \overline{u}.
\end{aligned}
\end{equation}

As a result, the prediction model becomes affine in the augmented decision variables and parameters, enabling a multiparametric quadratic programming formulation and the computation of an explicit piecewise-affine control law.

\subsection*{Resulting Quadratic Program}

Define the augmented parameter vector
\begin{equation}
\xi^\top :=
\begin{bmatrix}
x_0^\top & \chi_1^\top & \cdots & \chi_{n_\theta}^\top
\end{bmatrix},
\end{equation}
and the augmented decision variable
\begin{equation}
z^\top :=
\begin{bmatrix}
U^\top & V_1^\top & \cdots & V_{n_\theta}^\top
\end{bmatrix}.
\end{equation}

Then~\eqref{eq:approx_lifted_model_multi} can be written compactly as
\begin{equation}
X = \Phi\,\xi + \Gamma\,z,
\label{eq:X_aff_reg}
\end{equation}
with
\begin{equation*}
\Phi :=
\begin{bmatrix}
S_{x_0} & S_{\theta_1 x_0} & \cdots & S_{\theta_{n_\theta} x_0}
\end{bmatrix},
\end{equation*}
\begin{equation*}
\Gamma :=
\begin{bmatrix}
S_u & S_{\theta_1 u} & \cdots & S_{\theta_{n_\theta} u}
\end{bmatrix}.
\end{equation*}
Since only the true input $U$ is penalized, introduce the selector
\[
S_U :=
\begin{bmatrix}
I_{N n_u} & 0 & \cdots & 0
\end{bmatrix},
\]
so that $U = S_U z$.
To preserve convexity of the quadratic program, a small regularization term may be introduced to ensure that the Hessian is positive semidefinite \cite{bemporad2002}.
Substituting~\eqref{eq:X_aff_reg} into the cost function~\eqref{eq:condensed_mpc} yields
\begin{equation}
\min_{z}\;
\frac{1}{2} z^\top H z + f(\xi)^\top z
\qquad\text{s.t.}\qquad
G z \le b + E\,\xi,
\label{eq:qp_reg}
\end{equation}
where
\begin{align}
H &= 2\left(\Gamma^\top \bar{Q} \Gamma + S_U^\top \bar{R} S_U\right),\\
f(\xi) &= 2\,\Gamma^\top \bar{Q} \Phi\,\xi.
\end{align}

The constant term $\xi^\top \Phi^\top \bar{Q} \Phi \xi$ is omitted since it does not affect the optimizer.
Per-stage polyhedral constraints
\[
x_k \in \mathcal{X} := \{x : F_x x \le f_x\}, \qquad
u_k \in \mathcal{U} := \{u : F_u u \le f_u\}
\]
are stacked to yield
\[
F_X X \le f_X, \qquad F_U U \le f_U,
\]
with $F_X := I_N \otimes F_x$ and $f_X := \mathbf{1}_N \otimes f_x$.
Using $X=\Phi\xi+\Gamma z$ and $U=S_U z$, the constraints become
\begin{align}
F_X \Gamma z &\le f_X - F_X \Phi \xi,\\
F_U S_U z &\le f_U.
\end{align}

Hence,
\begin{equation}
G := \begin{bmatrix} F_X \Gamma \\ F_U S_U \end{bmatrix},\qquad
b := \begin{bmatrix} f_X \\ f_U \end{bmatrix},\qquad
E := \begin{bmatrix} -F_X \Phi \\ 0 \end{bmatrix}.
\label{eq:constraints_reg}
\end{equation}

Finally, the McCormick envelope constraints~\eqref{eq:mccormick} are added elementwise over the prediction horizon.

\subsection{Method II: Local explicit MPC for parametric dependencies} 

Starting from the affine decomposition introduced in \eqref{Parametric decomposition}, our objective is to preserve the QP structure and to obtain a control law that depends explicitly and linearly on~$\theta$. We approximate these matrix powers through a first–order local expansion using the Fr\'echet derivative \cite{higham2008}.

\subsection*{Fr\'echet-based Local Approximation}

For the matrix map $f(M)=M^p$, the Fr\'echet derivative at $\tilde{A}$ applied to a perturbation $\Delta A$ is
\begin{equation}
    L_f(\tilde{A})[\Delta A]
    = \sum_{r=0}^{p-1} \tilde{A}^r (\Delta A) \tilde{A}^{p-1-r}.
\end{equation}
Under the affine parameterization 
\[
    A(\theta) = \tilde{A} + \theta \Delta A,
\]
the power $A(\theta)^p$ admits the expansion
\begin{equation}
    A(\theta)^p
    = \tilde{A}^p + 
    \theta\,L_f(\tilde{A})[\Delta A]
    + O(\tilde{A},\theta\Delta A),
\end{equation}
where $O$ contains second and higher-order terms.  
Neglecting higher-order terms yields the approximation
\begin{equation}
\label{MatPow}
    A(\theta)^p \approx \tilde{A}^{\,p} + \theta\,L_f(\Delta A,p).
\end{equation}

\subsection*{Affine Structure of the Condensed Prediction Matrices}

Replacing \eqref{MatPow} in the condensed model in Definition~1 shows that the prediction matrices preserve an affine dependence on $\theta$:
\begin{equation}
\label{Affine_theta}
    S_x = \tilde{S}_x + \theta\,\Delta S_x,
    \qquad
    S_u = \tilde{S}_u + \theta\,\Delta S_u,
\end{equation}
where $\tilde{S}_x$ and $\tilde{S}_u$ denote the nominal condensed matrices evaluated at $\tilde{A}$, and
\begin{equation}
\bigl[\Delta S_x\bigr]_k
=
\begin{cases}
0, & k = 1,\\[2mm]
L_f(\Delta A,\,k-1), & k = 2,\dots,N .
\end{cases}
\end{equation}
The corresponding deviation in the input map is
\vspace{-0.4em}
\begin{equation}
\bigl[\Delta S_u\bigr]_{k,i}
=
\begin{cases}
L_f(\Delta A,\,k-1-i)\,B, & i \le k-1,\\[2mm]
0, & i > k-1 ,
\end{cases}
\end{equation}
\vspace{-1.8em}
\[
k,i = 0,\dots, N-1.
\vspace{-0.6em}
\]
These matrices preserve the structure shown in Definition~3 and incorporate the first-order parameter sensitivity.
We see that, in contrast to the condensed form in Definition~2, the affine structure proposed in \eqref{Affine_theta} still accounts for the parametric effects within the state evolution (through $\Delta S_x$) and also propagates its effects on all the inputs considered in the prediction (through $\Delta S_u$). Furthermore, this local approximation does not require the estimation of the  $E$ matrix required to propagate the artificial exogenous input introduced in the affine parametric representation \cite{diangelakis2017}.

\subsection*{Resulting Parameter-Dependent Quadratic Program}
Substituting \eqref{Affine_theta} into the condensed cost yields the quadratic expression:
\begin{equation}
     \textbf{U}^\top H(\theta)  \textbf{U} + 2 f(x_0, \theta)^\top  \textbf{U} + c,
\end{equation}
with
\vspace{-0.8em}
\begin{align*}
H &= \tilde{S}_u^\top \bar{Q} \tilde{S}_u + R  + \theta\!\left(
        \tilde{S}_u^\top \bar{Q} \Delta S_u
        + \Delta S_u^\top \bar{Q} \tilde{S}_u
        \right) \\
  &\quad + \theta^2 \Delta S_u^\top \bar{Q} \Delta S_u .
\end{align*}
\vspace{-1.8em}
\begin{align*}
f &= \tilde{S}_u^\top \bar{Q} \tilde{S}_x x_0 + \theta\!\left(
        \tilde{S}_u^\top \bar{Q} \Delta S_x x_0
        + \Delta S_u^\top \bar{Q} \tilde{S}_x x_0
        \right) \\
  &\quad + \theta^2 \Delta S_u^\top \bar{Q} \Delta S_x x_0 .
\end{align*}
\vspace{-1.8em}
\begin{align*}
c &= x_0^\top \tilde{S}_x^\top \bar{Q} \tilde{S}_x x_0 + 2\theta\, x_0^\top \Delta S_x^\top \bar{Q} \tilde{S}_x x_0 \\
  &\quad + \theta^2 x_0^\top \Delta S_x^\top \bar{Q} \Delta S_x x_0 .
\end{align*}

These expressions make the influence of the parameter $\theta$ on the quadratic cost components explicit.  
If a purely linear parameter dependence is desired, the second-order terms may be dropped.

\subsection*{First-Order Explicit Solution}
As a representative example, consider the unconstrained formulation together with the first-order truncation of the parameter-dependent cost. 
\[
    H(\theta) \approx H_0 + \theta\,\Delta H,
    \qquad
    f(\theta) \approx f_0 + \theta\,\Delta f,
\]
In this reduced setting, the optimal control sequence can be obtained using the following first-order inverse expansion
\begin{equation}
\label{invh_ap}
    H(\theta)^{-1}
    \approx H_0^{-1} - H_0^{-1}(\theta \Delta H)H_0^{-1}.
\end{equation}
Based on this approximation, the optimizer becomes
\begin{align}
U_0^*(\theta) = - H(\theta)^{-1} f(\theta), \\
    U_0^*(\theta)  \approx 
    U_0^* + \theta \,H_0^{-1}\big(\Delta H\,U_0^* - \Delta f\big)
    + \mathcal{O}(\theta^2),
\end{align}
where $U_0^* = -H_0^{-1} f_0$ is the solution using the nominal parameter.  
This expression explicitly shows how the optimal sequence varies with~$\theta$ and how the solution deviates from the nominal one when no constraint is active. Finally, when the system has state constraints, it is also necessary to approximate their condensed form to solve the quadratic program. Depending on the system, truncating the high-order terms can also alter the critical regions of the explicit MPC solution~\cite{rojas2026}. Therefore, it is important to carefully evaluate the validity region of the proposed model in terms of the parameters considered.

\section{CASE STUDIES}
\label{sec:Cases}
We demonstrate the methods on a thermal and a mechanical system using an MPC formulation equivalent to the feedback structure with preview proposed in \cite{nash2021,Theunissen2019Regionless}. The cost function penalizes tracking error and input effort, while initial-state and input bounds are imposed to reflect the physical limits of the system. We employ a small prediction horizon $N$ to keep the examples computationally tractable. Furthermore, to isolate the effects of the approximations, we adopt a nominal model and avoid additional terms, such as disturbance models and terminal ingredients, which can be incorporated into the proposed methods. The parameters used for the model and the control formulation are given in Table I.
\vspace{-2mm}
\begin{table}[h!]
\centering
\caption{Model parameters for case studies}
\label{tab:model_parameters}
\begin{tabular}{llll}
\hline
\textbf{Parameter} & \textbf{Description} & \textbf{Value} & \textbf{Unit} \\
\hline
\multicolumn{4}{c}{\textit{Mass--Spring--Damper System}} \\
\hline
$m$        & Mass of the moving body     & $2$            & \si{kg} \\
$c$        & Damping coefficient         & $3.109$        & \si{N.s/m} \\
$k_e$      & Spring stiffness            & $[500,\,2000]$  & \si{N/m} \\
$k_u$      & Input (actuator) gain       & $3.717$        & \si{N/V} \\
$z$        & Displacement of the mass    & $[-0.1,\,0.1]$  & \si{m} \\
$\dot{z}$  & Velocity of the mass        & $[-0.5,\,0.5]$  & \si{m/s} \\
$u(t)$     & Control input               & $[-5,\,5]$      & \si{V} \\
\hline
\multicolumn{4}{c}{\textit{Heat Exchanger System}} \\
\hline
$V$              & Fluid volume                    & $1000$        & \si{L} \\
$C$              & Heat capacity                    & $1$           & \si{kJ/(L.\celsius)} \\
$T_h$            & Hot outflow temperature          & $[45,\,70]$    & \si{\celsius} \\
$T_c$            & Cold outflow temperature         & $[25,\,45]$    & \si{\celsius} \\
$T_h^{\text{in}}$& Hot inflow temperature           & $[60,\,80]$    & \si{\celsius} \\
$T_c^{\text{in}}$& Cold inflow temperature          & $25$          & \si{\celsius} \\
$F_h$            & Flow rate of the hot fluid       & $2.5$         & \si{L/s} \\
$F_c$            & Flow rate of the cold fluid      & $2$           & \si{L/s} \\
$U$              & Heat transfer coefficient        & $1$           & \si{kJ/(s.m^2.\celsius)} \\
$A_h$            & Heat transfer area               & $[2,\,5]$      & \si{m^2} \\
\hline
\end{tabular}
\end{table}

\subsection{HEX: Heat exchanger}

We first consider a simple heat exchanger with a parametric dependence on
the heat transfer area, based on the study presented in~\cite{patilas2021}. The hot
fluid flows through a counter-flow heat exchanger with volumetric flow rate
$F_h$, entering at temperature $T_h^{\text{in}}$ and leaving at temperature
$T_h$. The cold fluid enters the heat exchanger at temperature
$T_c^{\text{in}}$ with volumetric flow rate $F_c$ and exits at temperature $T_c$.

\paragraph{Continuous-time model}
The lumped-parameter energy balance equations are given by
\begin{equation}
\begin{split}
    V \frac{d T_h}{dt} & = F_h \left( T_h^{\text{in}} - T_h \right)
    - \frac{U A_h}{C_r}(T_h - T_c), \\
    V \frac{d T_c}{dt} & = F_c \left( T_c^{\text{in}} - T_c \right)
    + \frac{U A_h}{C_r}(T_h - T_c).
\end{split}
\label{eq:HEXODE}
\end{equation}

In this setting, the inlet temperature of the hot stream,
$T_h^{\text{in}}$, is treated as the manipulated variable, while the cold-side outlet temperature $T_c$ is selected as the controlled output. The control objective is to track a desired reference trajectory for $T_c$ while respecting the system dynamics.

\paragraph{Affine parametric representation}

As discussed in Section~\ref{Parametric representation}, we consider a parametric discrete-time model of the HEX and include the effect of the constant exogenous input as a measured disturbance in the prediction of the systems dynamics:
\begin{equation}
x_{k+1} = A(\theta)\,x_k + B\,u_k + E\,d_k,
\qquad
y_k = Cx_k + Du_k ,
\end{equation}
where the state, input, and disturbance are defined as
\[
x_k := \begin{bmatrix} T_{h,k} \\ T_{c,k} \end{bmatrix}, 
\qquad
u_k := T^{\text{in}}_{h,k},
\qquad
d_k := T^{\text{in}}_{c,k}.
\]

The parameter-dependent heat transfer coefficient is defined as
\[
\alpha(\theta) := \frac{U A_h(\theta)}{C_r V},
\]
where $A_h(\theta)$ denotes the effective heat transfer area, $U$ is the overall
heat transfer coefficient, $C_r$ is the thermal capacitance, and $V$ is the fluid
volume.

Starting from \eqref{eq:HEXODE}, the system can be written in state-space form as
\begin{equation}
\dot{x}(t) = A_c(\theta)\,x(t) + B_c\,u(t) + E_c\,d(t),
\end{equation}
with
\begin{equation}
\label{AcHEXMatrix}
A_c(\theta) =
\begin{bmatrix}
-\dfrac{F_h}{V} - \alpha(\theta) & \alpha(\theta) \\[8pt]
\alpha(\theta) & -\dfrac{F_c}{V} - \alpha(\theta)
\end{bmatrix},
\end{equation}
\begin{equation}
B_c =
\begin{bmatrix}
\dfrac{F_h}{V} \\[4pt]
0
\end{bmatrix},
\qquad
E_c =
\begin{bmatrix}
0 \\[4pt]
\dfrac{F_c}{V}
\end{bmatrix}.
\end{equation}

The heat transfer area is assumed to vary within known bounds, $A_h \in [A_{\min}, A_{\max}]$, and is parameterized affinely as
\begin{equation}
A_h(\theta) = A_{\min} + \theta\,(A_{\max} - A_{\min}),
\qquad \theta \in [0,1].
\end{equation}

Using a forward Euler discretization with sampling time $T_s$, the resulting discrete-time state matrix can be expressed in affine form with respect to $\theta$ as in \eqref{Parametric decomposition}, with
\[
\tilde{A}_c := A_c(A_{\min}),
\qquad
\Delta A_c := A_c(A_{\max}) - A_c(A_{\min}).
\]
The discrete-time input and disturbance matrices are given by
\[
B := T_s B_c,
\qquad
E := T_s E_c.
\]

\subsection{MSD: Mass--spring--damper system}

As a second case study, we consider a mass--spring--damper system whose
parameters $(m, k_e, c)$ are taken from a real experimental setup
\cite{nash2021}. To obtain an analytical linear model, friction effects are
neglected in this initial study.

\paragraph{Continuous-time model}

The system dynamics are described by
\begin{equation}
    m\ddot z + c(\theta)\dot z + k_e z = k_u u(t),
    \qquad
    x = \begin{bmatrix} z \\ \dot z \end{bmatrix},
\end{equation}
where $z$ denotes the position of the mass, $\dot{z}$ its velocity, and $u(t)$ is the manipulated variable.

\paragraph{Affine parametric representation}

Following a similar approach to the previous case study, the discrete-time
parametric representation of the system is given by \eqref{GeneralSysDef} 
with
\begin{equation}
    A(\theta) =
    \begin{bmatrix}
        1 & T_s \\
        -\tfrac{k_e(\theta)}{m} T_s &
        1 - \tfrac{c}{m} T_s
    \end{bmatrix},
    \qquad
    B = 
    \begin{bmatrix}
        0 \\[2pt]
        \tfrac{T_s k_u}{m}
    \end{bmatrix}.
\end{equation}
In this case, the design parameter is the stiffness of the spring $ k_e(\theta)$ with limits defined in Table I. 

\section{RESULTS}
\label{sec:Results}
Using the case studies described in Section \ref{sec:Cases}, we compare the results obtained from the exact analytical solution implemented in YALMIP \cite{Lofberg2004} with those  obtained using the proposed methods described in Section \ref{sec:Methods}. The quantitative errors of the two methods with respect to the exact solution are presented in Table~\ref{tab:error_indices}.
\begin{table}[!ht]
\caption{Approximation error metrics for case studies}
\label{tab:error_indices}
\centering
\renewcommand{\arraystretch}{1.15}
\setlength{\tabcolsep}{2.0pt}

\begin{tabular}{c c c c c c c}
\hline
& \multicolumn{2}{c}{\textbf{RMSE}} 
& \multicolumn{2}{c}{\textbf{MAE}}
& \multicolumn{2}{c}{\textbf{NRMSE}} \\
\textbf{$\theta$}
& \textbf{Method I} & \textbf{Method II}
& \textbf{Method I} & \textbf{Method II}
& \textbf{Method I} & \textbf{Method II} \\
\hline
\multicolumn{7}{c}{\textit{Mass--Spring--Damper System}} \\
\hline
0.5
& 1.98e-4 & 5.13e-5
& 1.10e-3 & 4.60e-4
& 3.70e-2 & 9.60e-3 \\

1
& 3.89e-5 & 6.06e-5
& 2.42e-4 & 4.44e-4
& 7.40e-3 & 1.16e-2 \\
\hline

\multicolumn{7}{c}{\textit{Heat Exchanger System}} \\
\hline
0.5
& 4.43e-2 & 1.31e-1
& 7.52e-2 & 2.22e-1
& 1.10e-3 & 3.10e-3 \\

1
& 2.08e-2 & 2.58e-1
& 3.02e-2 & 4.65e-1
& 4.92e-4 & 6.10e-3 \\
\hline
\end{tabular}
\end{table}

\subsection{MSD numerical results}
For the MSD, we define a sampling time $T_s=0.01s$ and a horizon $N=4$. The error is penalized with $Q = 1000I$, and the input with $R=0.0001.$  The input and the initial states are constrained according to the values in Table I. The
results for $\theta=0.5 \ (k_e=1250)$ and $\theta=
1 \ (k_e=2000)$ are shown in Figures 1 and 2, respectively. In this case, we observe that both methods provide good quantitative and qualitative results, independent of the $\theta$ used. There are two reasons for this behavior. First, based on the controller response, we see that the system does not seem to be sensitive enough to different values of $k_e$. Second, the matrix $\Delta A$ obtained from the affine form \eqref{Parametric decomposition} has a single nonzero entry, located off the diagonal. This makes the higher-order terms in $A(\theta)^p$ negligible.

\begin{figure}[h]
\centering
\includegraphics[width=1.0\columnwidth]{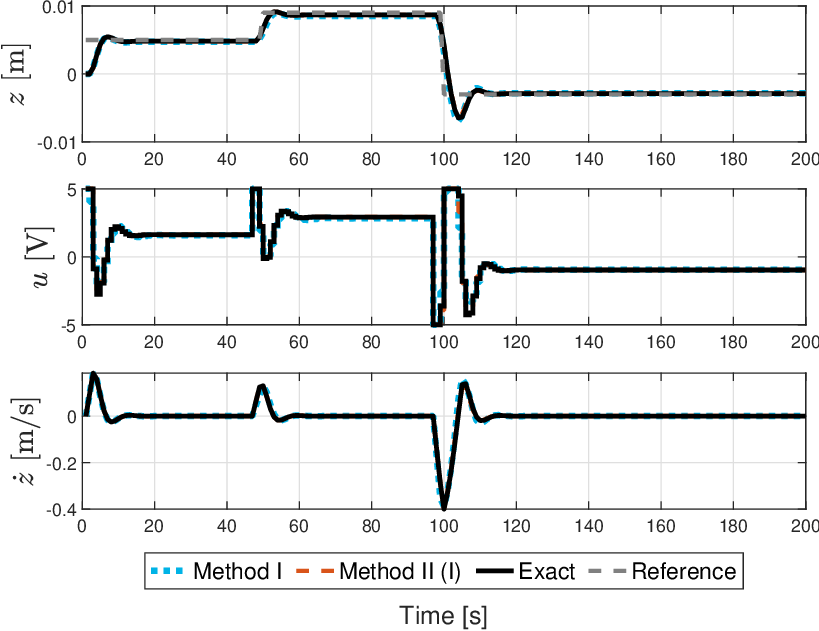} 
\caption{MSD: Exact vs Method I and Method II results for $\theta = 0.5.$}
\label{fig2}
\end{figure}
\begin{figure}[!h]
\centering
\includegraphics[width=1.0\columnwidth]{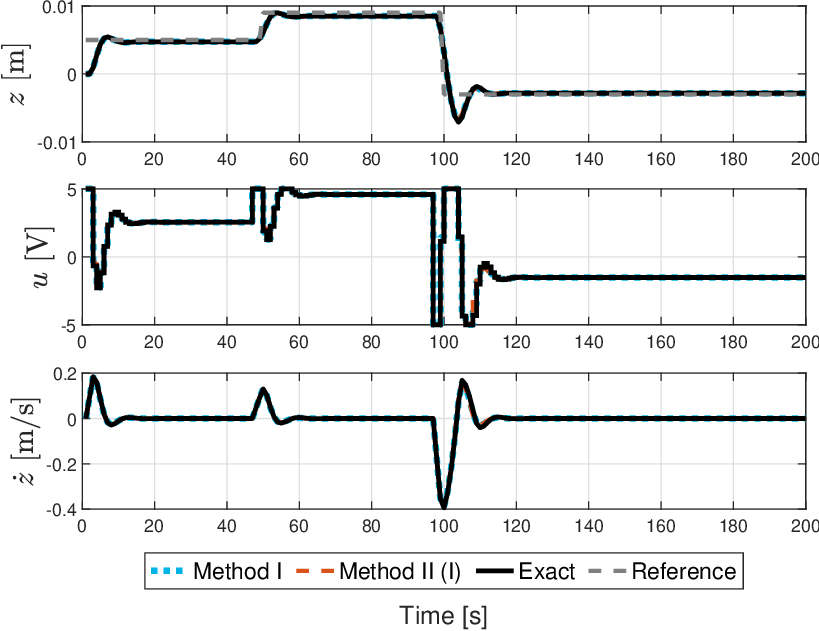} 
\caption{MSD: Exact vs Method I and Method II results for $\theta = 1.$}
\label{fig2}
\end{figure}

\subsection{HEX numerical results}
For the HEX, we consider $T_s = 10s$, $N= 4$, $Q = 100I$ and $R=0.001.$  The input and the initial state are constrained by the limits in Table I. For simplicity, the prediction and control horizons are the same, but input blocking or rate-of-change constraints could also be considered. Since Method II is based on an approximate inversion, we assume that there is a smooth change in every step of the controller, and we use the previous region as the controller’s fallback option.
The results for $\theta = 0.5 \ (A=3.5)$ and $\theta = 1 \ (A=5)$ are presented in Figures 3 and 4, respectively. We noticed that for $ A_h = 5 \ (\theta = 1$), the system is unable to follow the reference commanded at $t=400$, but this has to do with the physics of the HEX, not with the controllers themselves. We see that for $\theta = 0.5$ the response obtained with Method I follows the analytic solution almost exactly, while the solution obtained via Method II slightly deviates. In contrast, when $\theta = 1$, we observe some ripples in the output provided with Method II. Furthermore, the controller strategy computed by Method II has a bang-bang behavior (see Fig. 4). Based on this observation, we also implemented Method II without the inverse approximation \eqref{invh_ap}. We found that without the inclusion of the approximate inverse (Method II (NI)), the response of the system does not have the aforementioned bang-bang behavior. Regarding Method I, the inclusion of additional parameters results in a substantial increase in the number of regions compared to Method II. In the HEX case, Method I and Method II yield 10 and 704 regions, respectively, whereas in the MSD case the number of regions are 20 and 732. Despite this increase in complexity, Method I produces predictions that are numerically more stable and closer to the exact solution, indicating a trade-off between computational complexity and approximation performance.

\begin{figure}[h]
\centering
\includegraphics[width=1.0\columnwidth]{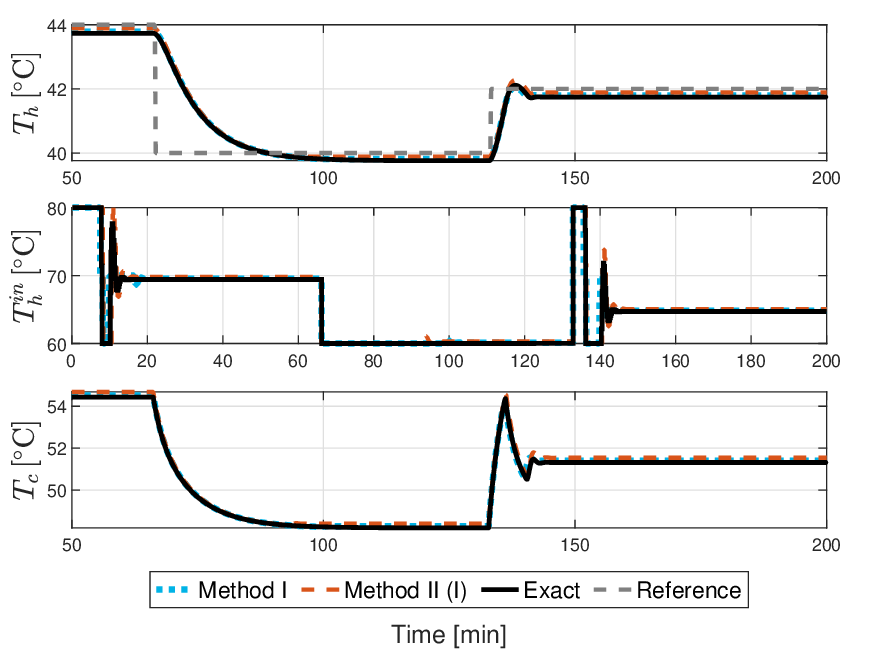} 
\caption{HEX: Exact vs Method II results for $\theta = 0.5.$}
\label{fig2}
\end{figure}

\begin{figure}[!ht]
\centering
\includegraphics[width=1.0\columnwidth]{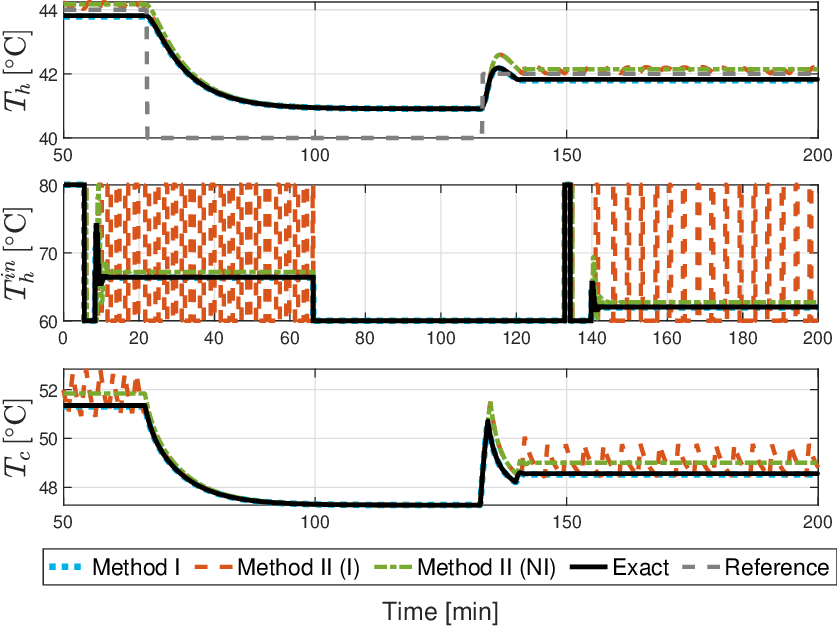} 
\caption{HEX: Exact vs Method II results for $\theta = 1.$}
\label{fig2}
\end{figure}
\vspace{-5mm}
\section{CONCLUSIONS}
\label{sec:Conclusion}
This paper addressed the problem of eMPC formulation for parameter dependent linear systems. Two approximation methods were proposed to mitigate the resulting nonlinear terms in the mpQP problem. Both methods rely on a consistent state-space representation and employ established mathematical approximation techniques. The case studies demonstrated that the proposed methods provide a good approximation depending on the sensitivity of model to parameter and for certain values of $\theta$. The future work will consider the incorporation  of integral action in eMPC and application of control problems characterized by parameter-dependent dynamics, with particular emphasis on control co-design problems.

\addtolength{\textheight}{-12cm}   





\section*{ACKNOWLEDGMENT}

The authors acknowledge Dr. Valentina Breschi (TU Eindhoven) for the helpful discussions regarding the development of Method II and Dr. Cesar Augusto Perez (TU Eindhoven) for the helpful discussions regarding the development of Method I.



\end{document}